\documentclass[preprint,aps,floatfix]{revtex4-1}
\usepackage{hyperref}
\everymath={\displaystyle}
\usepackage{young}
\usepackage{rotating}
\usepackage{longtable}
\def\1{\mbox{l\hspace{-0.53em}1}}

\usepackage{multirow}
\newlength{\AccoHaut}

\begin{document}
\title{Relations between strong decay widths of the $P_c$ pentaquarks in the SU(4)  flavor-spin model}
\author{Fl. Stancu\thanks{e-mail : fstancu@ulg.ac.be}}
\address{Universit\'{e} de Li\`ege, Institut de Physique B.5, Sart Tilman,
B-4000 Li\`ege 1, Belgium}

\date{\today}
\everymath={\displaystyle}

\begin{abstract}

In a previous work we have studied the isospin 1/2 lowest   
positive and  negative parity states   of the
pentaquark $uudc\overline{c}$,  
in a constituent quark model with a linear confinement and an SU(4) flavor-spin hyperfine interaction
and we compared the results 
with the  $P^+_c(4312)$, $P^+_c(4440)$ and $P^+_c(4457)$ pentaquarks observed at LHCb in 2019.
Here we extend the previous work by calculating 
ratios of decay rates  of the $P_c$ pentaquarks to $J/\Psi$ and  $\eta_c$ 
and similarly the ratio of decay rates to $\Lambda_c {\bar D}^*$ and $\Lambda_c \bar D$.
Our predictions are based  on the
SU(4)$\times$SU(2) structure of compact pentaquarks. 

\end{abstract}

\maketitle

\vspace{1cm}

\section{Introduction}

The observation of the narrow structures  $P^+_c(4312)$, $P^+_c(4440)$ and $P^+_c(4457)$
in the $\Lambda^0_b \rightarrow J/\psi K^- p$ decay made by LHCb in 2019 and its interpretation as a pentaquark
with flavor content $uudc\overline{c}$ \cite{Aaij:2019vzc}
has stimulated considerable interest in further understanding the structure of these pentaquarks.

Although observed in the $J/\psi p$ channel,
the proximity of the mass of the $P^+_c(4312)$ to the $\Sigma^+_c \overline{D}^0$ threshold  (4318 MeV)
and of the masses of  $P^+_c(4440)$ and $P^+_c(4457)$ to the  $\Sigma^+_c \overline{D}^{*0}$ threshold (4460 MeV),
favored  their interpretation as molecular S-wave of the   $\Sigma^+_c + \overline{D}^0$ 
and $\Sigma^+_c + \overline{D}^{*0}$  systems respectively
 \cite{Guo:2019kdc,Guo:2019fdo,Xiao:2019mst,Xiao:2019aya,Lin:2019qiv,
Liu:2019tjn,Meng:2019ilv,Wu:2019rog,Valderrama:2019chc,Du:2019pij,Wang:2019spc,Xu:2020gjl,Chen:2020pac,
Fernandez-Ramirez:2019koa}.
In such an interpretation, 
the binding arises via meson exchanges between point particles and in the elastic channel 
all resonances acquire a negative parity. 

The spectrum of the $uudc\overline{c}$ pentaquark has also been analyzed in compact pentaquark models. 
An advantage with respect to molecular models is that they allow a classification of pentaquarks
into multiplets \cite{Ortiz-Pacheco:2018ccl}. A considerable amount of
studies are based on  the color-spin (CS) chromomagnetic interaction 
of the one gluon exchange model 
with quark/antiquark correlations \cite{Ali:2019npk} or without correlations, see, for example, Refs. \cite{Weng:2019ynv,Cheng:2019obk}.

In Ref. \cite{Stancu:2019qga} we have studied the spectrum of the $uudc\overline{c}$ pentaquark 
and in Ref. \cite{Stancu:2020paw} the spectrum of the isoscalar $udsc\overline{c}$ pentaquark
within a model based on the SU(4) flavor-spin (FS) hyperfine interaction 
obtained as an extension of 
the meson exchange model between quarks \cite{Glozman:1995fu,Glozman:1996wq}
to both light and heavy meson exchange. 
The  flavor-spin model  provides a good description of low-lying nonstrange and strange baryons
by correctly reproducing the  order of positive and negative parity states in contrast to models based on the hyperfine
color-spin chromomagnetic  interaction.

The  extension  to SU(4) has been made in the spirit of the phenomenological approach
of Ref. \cite{Glozman:1995xy} where, in addition to Goldstone bosons of 
the hidden approximate chiral symmetry of QCD,  the flavor exchange interaction was augmented  
by an additional exchange of $D$ mesons between $u, d$ and $c$ quarks and 
of $D_s$ mesons between $s$ and $c$ quarks.

The conclusion was that the lowest state of the $uudc\overline{c}$ pentaquark has negative parity
for the CS interaction and positive parity for the FS interaction.
The spin and parity of the narrow structures  $P^+_c(4312)$, $P^+_c(4440)$ and $P^+_c(4457)$ are presently unknown experimentally.

The parity of the
pentaquark is given by P$\ = {\left({-}\right)}^{{\ell\ +\ 1}}$, where  $\ell$ is the orbital angular momentum
of the excited system.
For the lowest positive parity states
one way is to  introduce an angular momentum  $\ell$ = 1 in the internal motion of the four-quark subsystem.
According to the Pauli principle, the four-quark subsystem  must be in a state of
orbital symmetry ${\left[{31}\right]}_{O}$. 
Although the kinetic energy of ${\left[{31}\right]}_{O}$
is higher than that of the totally symmetric ${\left[{4}\right]}_{O}$
state of negative parity, the flavor-spin interaction overcomes this excess and generates a lower 
eigenvalue for the   ${\left[{31}\right]}_{O}$  state  with an  $s^3p$ configuration  than for 
${\left[{4}\right]}_{O}$  with an $s^4$ configuration \cite{Stancu:2019qga}.

The present work aims at understanding the role of the flavor-spin structure of the wave functions of the $uudc\overline{c}$
pentaquarks  studied in Ref. \cite {Stancu:2019qga}
on some of their strong decay properties. We restrict our considerations to the lowest positive and 
negative parity states  $J^P = 1/2^+$ and $J^P = 1/2^-$, respectively.

The study of strong decay  properties of $P_c$  pentaquarks is a present challenge.
A list of meson-baryon systems into which the $P_c$ pentaquarks of positive or negative parity can decay was presented
in Ref. \cite{Burns:2015dwa}.
Some strong decay properties of the 2019  LHCb pentaquarks have been considered 
in the framework of the baryon-meson molecules scenario \cite{Voloshin:2019aut,Burns:2019iih}.
The decay widths of the LHCb pentaquark $P_c(4312)$ has also been analyzed, for example, in a chiral constituent quark
model containing both chromomagnetic and meson exchange interactions \cite{Dong:2020nwk}.

Our  work is similar in spirit to that of  Ref. \cite{Voloshin:2019aut} where the molecular scenario was used
to calculate ratios of rates of decays to various channels. Here we study the role of the flavor-spin structure 
of the pentaquark wave functions on the ratios of decays of the $uudc \bar c$ to $J/\psi$ and to $\eta_c$ and 
of the decays to  $\Lambda_c \bar D$ and to $\Lambda_c {\bar D}^*$. This is achieved by calculating the 
overlap between the flavor-spin part of the pentaquark wave function and the flavor-spin part of the decay channel 
wave function.

The paper is organized as follows. In the next section we reproduce the flavor-spin Hamiltonian model
generalized to SU(4)  \cite{Stancu:2019qga,Stancu:2020paw}. In Sec. \ref{wavefunction}
we describe the SU(4) symmetry structure of the lowest positive and negative parity states 
named $P_c(1/2^+)$ and $P_c(1/2^-)$ respectively. In Sec. \ref{widths} we calculate the overlap 
between the pentaquark and the decay channel flavor-spin wave functions from which we derive the ratio
of decay rates in a simple manner. The last section is devoted to conclusions. Appendix \ref{baryons}
is a remainder of the flavor states of baryons into which the pentaquark can  decay. Appendix \ref{flavorpart}
describes the flavor states of the pentaquark obtained in an SU(4) classification \cite{Ortiz-Pacheco:2018ccl}.
In Appendix \ref{spinpart} we derive the spin part of the pentaquark wave function and in Appendix \ref{negative}
we present some useful details of the flavor-spin wave function of negative parity states.

\section{The Hamiltonian}\label{Hamiltonian}
Here we closely follow the description of the model used to calculate the spectrum of the $uudc \bar c$
pentaquark in Ref. \cite{Stancu:2019qga}, extended to strange hidden charm pentaquarks in Ref. \cite{Stancu:2020paw}.        
The Hamiltonian in its general SU(4) form is 
\begin{eqnarray}
H &=& \sum_i m_i + \sum_i\frac{{\vec p}_{i}^2}{2m_i} 
- \frac {(\sum_i {\vec p}_{i})^2}{2\sum_i m_i} + \sum_{i<j} V_{\text{conf}}(r_{ij}) \nonumber\\
&+& \sum_{i<j} V_\chi(r_{ij}),
\label{ham}
\end{eqnarray}
with $m_i$ and ${\vec p}_{i}$ 
denoting the quark masses and momenta respectively
and $r_{ij}$ the distance between the interacting quarks or quark-antiquark $i$ and $j$. 
The Hamiltonian contains the internal kinetic energy and the linear confining interaction 
\begin{equation}
 V_{\text{conf}}(r_{ij}) = -\frac{3}{8}\lambda_{i}^{c}\cdot\lambda_{j}^{c} \, C\, r_{ij} \, .
\label{conf}
\end{equation}
The hyperfine part $V_\chi(r_{ij})$ has a flavor-spin structure extended to SU(4) in Ref. \cite{Stancu:2019qga}
which has the following form
\begin{eqnarray}
V_\chi(r_{ij})
&=&
\left\{\sum_{F=1}^3 V_{\pi}(r_{ij}) \lambda_i^F \lambda_j^F \right. +  \sum_{F=4}^7 V_{K}(r_{ij}) \lambda_i^F \lambda_j^F 
\nonumber \\
&+& \left.  V_{\eta}(r_{ij}) \lambda_i^8 \lambda_j^8 
+V_{\eta^{\prime}}(r_{ij}) \lambda_i^0 \lambda_j^0\right. 
\nonumber \\
&+& \left. \sum_{F=9}^{12} V_{D}(r_{ij}) \lambda_i^F \lambda_j^F\right.     
+ \left. \sum_{F=13}^{14} V_{D_s}(r_{ij}) \lambda_i^F \lambda_j^F \right.
\nonumber \\
&+& \left. V_{\eta_c}(r_{ij}) \lambda_i^{15} \lambda_j^{15} \right\}
\vec\sigma_i\cdot\vec\sigma_j, 
\label{VCHI}
\end{eqnarray}
\noindent
with the SU(4) generators $\lambda^F_i$ ($F$ = 1,2,...,15)
and
$\lambda^0_i = \sqrt{2/3}~{\bf 1}$, where $\bf 1$ is the $4\times4$ unit
matrix. 
After integration in the flavor space, the two-body matrix elements  
containing contributions due to light,  strange and charm quarks are 
\begin{equation}\label{twobody}
V_{ij} =
{\vec {\sigma}}_i\ \cdot {\vec {\sigma}}_j\, 
\left\{ \renewcommand{\arraystretch}{2}
\begin{array}{cl}
 V_{\pi} + \frac{1}{3} V^{uu}_{\eta} + \frac{1}{6} V^{uu}_{\eta_c},  &\hspace{0.3cm} \mbox{ $[2]_F, I = 1$} \\
 2 V_K - \frac{2}{3} V^{us}_{\eta}, ~~~ 2 V^{uc}_D- \frac{1}{2} V^{uc}_{\eta_c} & \hspace{0.3cm} \mbox{ $[2]_F, I = \frac{1}{2}$} \\
 2 V^{sc}_{D_s} - \frac{1}{2} V^{sc}_{\eta_c} & \hspace{0.3cm} \mbox{ $[2]_F, I = 0$} \\
 \frac{4}{3} V^{ss}_{\eta} + \frac{3}{2} V^{cc}_{\eta_c} & \hspace{0.3cm} \mbox{ $[2]_F, I = 0$} \\
-2 V^{sc}_{D_s} - \frac{1}{2} V^{sc}_{\eta_c} & \hspace{0.3cm} \mbox{ $[11]_F, I = 0$} \\
-2 V_K - \frac{2}{3} V^{us}_{\eta},~~~ - 2 V^{uc}_D - \frac{1}{2} V^{uc}_{\eta_c} &\hspace{0.3cm} \mbox{ $[11]_F, I = \frac{1}{2}$} \\
-3 V_{\pi} + \frac{1}{3} V^{uu}_{\eta} + \frac{1}{6} V^{uu}_{\eta_c},  &\hspace{0.3cm} \mbox{ $[11]_F, I = 0$}
\end{array} \right.
\end{equation}
\noindent
In Eqs. (\ref{twobody}) the pair  of quarks $ij$ is either in a symmetric [2]$_F$ or in an antisymmetric [11]$_F$ flavor state
and the isospin $I$ is defined by the quark content. The upper index of 
$V$ exhibits the flavor of the
two quarks interchanging a meson specified by the lower index.
Obviously, in every sum/difference of Eq. (\ref{twobody}) the upper index is the same for all terms.

Thus  the SU(4) version of
the interaction (\ref{VCHI})
contains $\gamma = \pi, K, \eta, D, D_s, \eta_c$ and $\eta '$
meson-exchange terms.  Every $V_{\gamma} (r_{ij})$ is
a sum of two distinct contributions: a Yukawa-type potential containing
the mass of the exchanged meson and a short-range contribution of opposite
sign, the role of which is crucial in baryon spectroscopy. 
For a given meson $\gamma$ the meson exchange potential is
\begin{eqnarray}\label{radialform}
V_\gamma (r) &=&
\frac{g_\gamma^2}{4\pi}\frac{1}{12m_i m_j}
\{\theta(r-r_0)\mu_\gamma^2\frac{e^{-\mu_\gamma r}}{ r} \nonumber\\
&-& \frac {4}{\sqrt {\pi}}
\alpha^3 \exp(-\alpha^2(r-r_0)^2)\}
\end{eqnarray}

In the calculations of the spectrum of  $uudc \bar c$ we used 
parameters of Ref. \cite{Glozman:1996wq} to which we added the $\mu_{D}$ 
mass and
the coupling constants $\frac{g_{Dq}^2}{4\pi}$. The contribution of  $V_{{\eta}_c}$ can be neglected. We have 
$$\frac{g_{\pi q}^2}{4\pi} = \frac{g_{\eta q}^2}{4\pi} =
\frac{g_{Dq}^2}{4\pi}=  0.67,\,\,
\frac{g_{\eta ' q}^2}{4\pi} = 1.206 , $$
$$r_0 = 0.43 \, \mbox{fm}, ~\alpha = 2.91 \, \mbox{fm}^{-1},~~
 C= 0.474 \, \mbox{fm}^{-2}, \, $$
$$ \mu_{\pi} = 139 \, \mbox{ MeV},~ \mu_{\eta} = 547 \,\mbox{ MeV},~
\mu_{\eta'} = 958 \, \mbox{ MeV},~ \mu_{D} = 1867 \, \mbox{ MeV},~. $$
The meson masses correspond to the experimental values from  the  Particle Data Group 
\cite{Zyla:2020zbs}.
The model  has previously been used to study the stability of
open flavor tetraquarks \cite{Pepin:1996id} and open flavor pentaquarks \cite{Genovese:1997tm}.
Accordingly,
for the quark masses $m_{u,d}$ and $m_c$   we take  
\begin{equation}\label{quarkmass}
  m_{u,d} = 340 \, \mbox{ MeV}, ~ m_c = 1350 \, \mbox{ MeV}.
\end{equation}
They were adjusted to satisfactorily reproduce the average mass ${\overline M} = (M + 3 M^*)/4$ = 2008 MeV of
$D$ mesons. 
Using the above parameters the calculated baryon masses
are  $m_N$ = 960 MeV, $m_{\Lambda_c}$ = 2180 MeV and $m_{\Sigma_c}$ = 2434 MeV  \cite{Stancu:2019qga}.

\section{The pentaquark wave function}\label{wavefunction}

The pentaquarks under discussion are denoted by $P_c(1/2^+)$ and $P_c(1/2^-)$, which are the lowest  positive and negative parity 
states of the Hamiltonian introduced in Sec. \ref{Hamiltonian}. 
The parity of a pentaquark is $P = (-1)^{\ell+1}$.
The pentaquark wave functions 
showing the symmetry structure  of the four-quark subsystem
in terms of the orbital (O), color (C), flavor (F) and spin (S) degrees of freedom are
\begin{equation}\label{Pcplus}
P_c(1/2^+) = \ 
\left({{\left[{31}\right]}_{O}{\left[{211}\right]}_{C}{\left[{{1}^{4}}\right]}_{OC}\
;\
{\left[{22}\right]}_{F}{\left[{22}\right]}_{S}{\left[{4}\right]}_{FS}}\right) \phi(\bar c)~,
\end{equation}
\begin{equation}\label{Pcminus}
P_c(1/2^-) = \ 
\left({{\left[{4}\right]}_{O}{\left[{211}\right]}_{C}{\left[{211}\right]}_{OC}\
;\
{\left[{211}\right]}_{F}{\left[{22}\right]}_{S}{\left[{31}\right]}_{FS}}\right) \phi(\bar c)~.
\end{equation}
The $\phi(\bar c) $ is the wave function of the antiquark with the same degrees of freedom. The antiquark  
is coupled to the $q^4$ subsystem in the color, flavor and spin spaces. The  coupling in the flavor space to
a definite flavor symmetry of $q^4 \bar q$ is described in Appendix \ref{flavorpart} and  the coupling to 
a total spin $S$ = 1/2 for both $P_c(1/2^+)$ and $P_c(1/2^-)$ is presented in  Appendix \ref{spinpart}.   
The total angular momentum is $\vec{J} = \vec{L} + \vec{S}$, so that for $\ell = 1$ the pentaquark  $P_c(1/2^+)$ 
can have quantum numbers $J^P = 1/2^+$ or $ 3/2^+$.
The two states are degenerate so that 
the latter quantum number can be omitted in  the discussion. 
The $q^4 \bar q $ is in a color singlet state. 
The orbital parts of $P_c(1/2^+)$ and $P_c(1/2^-)$ are described in detail in Ref. \cite {Stancu:2019qga}. 
They are  defined in terms of the internal coordinates of five particles and are translationally invariant 
(no center of mass motion).
In the following the expressions
of the orbital wave functions are not necessary.

%
From Eqs. (\ref{Pcplus}) and (\ref{Pcminus})
one can see that the $q^4$ subsystem in $P_c(1/2^+)$ is in a symmetric flavor-spin state
and that in   $P_c(1/2^-)$  the subsystem $q^4$ is in an antisymmetric color-flavor-spin state. Together  
with the orbital part a totally antisymmetric state is obtained in both cases.

The pentaquark wave function results from the coupling of $q^4$ to $\bar q$ in the color, flavor and spin spaces.
The expression of the flavor-spin part of the wave function
of  $P_c(1/2^+)$  
is obtained from the  Clebsch-Gordan coefficients \cite{Stancu:1991rc}
of the inner product  $[22]_F \otimes [22]_S \rightarrow [4]_{FS}$ because 
after the coupling the flavor-spin wave function  of $q^4$  
remains symmetric under the permutation group $S_4$. 
Then  
the flavor-spin wave function of  $P_c(1/2^+)$ has the following form 
\begin{equation}\label{FSplus} 
(\phi \chi)^5_E  = \frac{1}{\sqrt{2}} (\phi^{\rho}_E \chi^{\rho}_{SM} + \phi^{\lambda}_E \chi^{\lambda}_{SM}),   
\end{equation}
in terms of flavor and spin states defined in Appendices \ref{flavorpart}  and \ref{spinpart}  respectively.
The upper indices $\rho$ and $\lambda$ indicate that the flavor or spin state is  antisymmetric and symmetric 
respectively under interchange of the particles 1 and 2, like for the nucleon. 
Here and below
the upper index 5 stands for five particles, i. e. the pentaquark.

In a similar way,  
the color-flavor-spin wave function  of $P_c(1/2^-)$
can be obtained from the Clebsch-Gordan coefficients of the inner product 
$[211]_C \otimes [31]_{FS} \rightarrow [1^4]_{CFS}$ inasmuch as the flavor-spin-color wave function  of $q^4$ 
remains antisymmetric under $S_4$. One has
 \begin{equation}\label{FSminus} 
(\phi \chi C)^5_{F_{1}} = \frac{1}{\sqrt{3}} [ (\phi_{F_{1}} \chi_{SM})_1 C_1 - (\phi_{F_{1}} \chi_{SM})_2 C_2 +  (\phi_{F_{1}} \chi_{SM})_3 C_3],
\end{equation}
where $C_i$ are the three independent basis vectors of the irreducible representation $[222]_C$ describing 
the color singlet $q^4 \bar q$ system, which results from the SU(3)-color  direct product decomposition
$[211] \otimes [11]$. 
These are 
\begin{equation} 
C_1 = 
\raisebox{-15.0pt}{\mbox{\begin{Young}
1 & 2 \cr
3 & 5 \cr
4 & 6 \cr
\end{Young}}}\\[2.1ex]~, ~~~~~
C_2 = 
\raisebox{-15.0pt}{\mbox{\begin{Young}
1 & 3 \cr
2 & 5 \cr
4 & 6 \cr
\end{Young}}}\\[2.1ex]~, ~~~~~
C_3 = 
\raisebox{-15.0pt}{\mbox{\begin{Young}
1 & 4 \cr
2 & 5 \cr
3 & 6 \cr
\end{Young}}}\\[2.1ex]~.
\end{equation}
The only one which gives a non-vanishing overlap with the color wave function of the exit channel 
is $C_3$.  Its explicit form is not needed in the calculation of $O^{FS}$.
The corresponding flavor-spin state  $(\phi_{F_{1}} \chi_{SM})_3$ with $S$ = 1/2
is written explicitly in Appendix \ref{negative} in terms of flavor and spin parts.

The exit channel is formed of a baryon $B$ and a meson $M$. For $S$-waves decays, the corresponding flavor-spin state 
is defined by
\begin{equation}\label{baryonmeson}
|B M \rangle =  \frac{1}{\sqrt{2}} [(\phi^{\rho}_B ~{c \bar c}) ~(\chi^{\rho}_B \chi_{M})
+ (\phi^{\lambda}_B ~{c \bar c})~(\chi^{\lambda}_B \chi_{M})],  
\end{equation}
where $\phi^{\rho}_B$ and $\phi^{\lambda}_B$ are the usual SU(3) octet baryon states of mixed symmetry $[21]$,
The corresponding spin states of the baryon are $\chi^{\rho}_B$ and $\chi^{\lambda}_B$
and   $ \chi_{M}$
is the spin state of the meson. The baryon and the meson spins are coupled together to a value equal to that of the pentaquark spin.


In Ref. \cite{Stancu:2019qga} we have studied the mass spectrum of the   $uudc\overline{c}$ pentaquark
for several states including those here named $P_c(1/2^+)$ and $P_c(1/2^-)$. The calculated masses are in the range of the observed 
narrow structures $P^+_c(4312)$, $P^+_c(4440)$ and $P^+_c(4457)$. The SU(4) flavor-spin model described in Sec. \ref{Hamiltonian}
gives 4273 MeV   for the mass of $P_c(1/2^+)$,  see Table \ref{overlap}, and
it can tentatively be assigned to $P^+_c(4312)$.

The lowest negative parity state $P_c(1/2^-)$ has a mass of 4487 MeV, indicated in Table \ref{overlap}, close to that of the $P^+_c(4440)$ and $P^+_c(4457)$
resonances. Therefore it supports the quantum number $J^P = {\frac{1}{2}}^-$ for one of them. 
In discussing the decay widths we shall take into account these assignments.

\section{Decay widths}\label{widths}

The decay width of a pentaquark resonance into a baryon + meson  is proportional to the 
square of  the transition amplitude matrix element  between the initial  and  final states 
(see $\it e.g.$ Ref. \cite{Gasiorowicz}). 
In the present model we suppose that the transition amplitude matrix element  can be written as
a product of two factors, one containing the orbital-color degrees of freedom and the other
the flavor-spin degrees of freedom. In a simple estimate each  factor is proportional to the
overlap between the initial pentaquark and the  baryon + meson channel wave functions in the corresponding 
degrees of freedom.  In the following we shall denote them by $O^{OC}$ and $O^{FS}$ respectively.
In the fall-apart mode considered here  we think that $O^{FS}$ is a good approximation to the
flavor-spin part of the transition amplitude matrix element because it contains the basic 
quark interchange  operator through the antisymmetrization  of the four quark subsystem wave functions,
as defined by Eqs. (\ref{Pcplus}) and (\ref{Pcminus}).
The orbital-color part of the transition amplitude deserves a special discussion.
Presently it is not needed because we are interested in ratios of decay rates o
f channels with the same symmetry in the
 flavor-spin space, where the  orbital-color part simplifies.

Using  Eqs. (\ref{FSplus}) and (\ref{baryonmeson}) the flavor-spin overlap of $P_c(1/2^+)$ can be written as
\begin{eqnarray}\label{OFSPLUS}
O^{FS}& = & \langle \phi \chi)^5_E |  B M  \rangle \nonumber \\
      & = & \frac{1}{2} ( \langle \phi^{\rho}_E | \phi^{\rho}_N c {\bar c} \rangle ~\langle \chi^{\rho}_5 | \chi^{\rho}_N \chi_{SM} \rangle 
        +  \langle \phi^{\lambda}_E | \phi^{\lambda}_N c {\bar c} \rangle ~\langle \chi^{\lambda}_5 | \chi^{\lambda}_N \chi_{SM} \rangle )~. 
\end{eqnarray}
From Eqs. (\ref{Erho})  and (\ref{Elambda}) associated to the pentaquark $uudc \bar c$ flavor wave functions
and the spin wave functions of Appendix \ref{spinpart} we have obtained the overlap $O^{FS}$ 
for the decay channels where $B = p$ and  $B = \Lambda_c$, as indicated in Table  \ref{overlap}. 
One can see that the largest overlap  is 0.5303 which corresponds to  $\Lambda_c+{\bar D}^*$.

Neglecting kinematical differences  and other common factors 
one can easily find that the ratio of the decay rates for $B = p$ is 
\begin{equation}\label{pdecaypositive} 
\frac{\Gamma(P_c(1/2^+) \rightarrow \eta_c~ p)}{\Gamma(P_c(1/2^+) \rightarrow J/\psi~ p)} = \frac{1}{3}~,
\end{equation}
and for $B = \Lambda_c$ the ratio is
\begin{equation}\label{lambdadecay} 
\frac{\Gamma(P_c(1/2^+) \rightarrow \Lambda_c \bar D)}{\Gamma(P_c(1/2^+) \rightarrow \Lambda_c {\bar D}^*)} = \frac{1}{3}.
\end{equation}
The above ratios are equal for the following reason. The flavor part contribution to $O^{FS}$ cancels out  in the numerator 
and the denominator when the scalar and vector mesons  have the same flavor content. 
This is obviously the case for $\bar D$ and  ${\bar D}^*$. 

The quark content of $\eta_c$ is not  
known experimentally. Here we have assumed that it has the same content as $J/\psi$, compatible with
an ideal mixing.  
The value of 1/3 corresponds to the square of the 
ratio of the mixing coefficients in the spin wave function of the pentaquark as given by Eq. (\ref{pentaspin}).

If $P_c(1/2^+)$ is identified with the observed   $P^+_c(4312)$ resonance we can make a comparison
with the results of Ref. \cite{Voloshin:2019aut} based on the molecular scenario. We note  
that the ratio (\ref{pdecaypositive}) is the inverse of the value predicted from the molecular scenario
when considered as a $J^P = 1/2^-$ $\Sigma_c {\bar D}^*$ molecule.

After integration in the color space, the overlap $O^{FS}$
of the lowest negative parity pentaquark $P_c(1/2^-)$ described by the wave function (\ref{FSminus})
becomes 
\begin{equation}
O^{FS} = \frac{1}{\sqrt{3}} \langle (\phi_{F_{1}} \chi_{SM})_3 | BM \rangle.
\end{equation}
The presence of the factor $1/\sqrt{3}$ is due to the norm of (\ref{FSminus}) and
as mentioned above only the third term of this wave function contributes to fall-apart decays.
The integration in the flavor-spin space is made using Appendix \ref{negative}. 
The largest overlap is  0.3062  and corresponds to the $p+J/\Psi$ channel for the lowest negative parity pentaquark.

From the resulting expression of the overlap we find that the ratio of the decay rates for $B = p$ is 
\begin{equation}\label{pdecaynegative} 
\frac{\Gamma(P_c(1/2^-) \rightarrow \eta_c~ p)}{\Gamma(P_c(1/2^-) \rightarrow J/\psi~ p)} = \frac{1}{3}~,
\end{equation}
i. e. the same as for $ P_c(1/2^+)$, for the same reason.   
In Ref. \cite{Voloshin:2019aut} the pentaquark $P_{c1}$ with $J^P = 1/2^-$ is assigned to  the observed $P^+_c(4440)$ resonances.
The ratio of decay rates to $\eta_c~ p$ and $J/\psi~ p$ is at variance with our result.

\begin{table}
\caption{\label{overlap}Lowest positive and negative parity  $uudc \bar c$ pentaquarks
of quantum numbers $S$ and  $J^P$ and symmetry structure defined in (\ref{Pcplus}) and (\ref{Pcminus}).
Column 1 gives the name, column 2 the spin, column 3 the total angular momentum and parity, column 4 the mass 
calculated in Ref. \cite{Stancu:2019qga} and columns 5-8 
the value of the overlap $O^{FS}$ for the indicated decay channel.}
\begin{tabular}{cccccccccccc}
\hline
Pentaquark & \hspace{3mm}$S$ & $J^P$ & \hspace{2mm} Mass & \multicolumn{5}{c}{Decay channel}   \\
&\hspace{6mm} & \hspace{6mm} & \hspace{2mm} (MeV)  & \hspace{2mm} $p+J/\psi$ & \hspace{2mm} $p+\eta_c$ & \hspace{2mm} $\Lambda_c+{\bar D}^*$ 
& \hspace{2mm} $\Lambda_c+{\bar D}$ & \hspace{2mm}  \\[1.1ex]
\hline 
$P_c(1/2^+)$ & \hspace{1mm} $\frac{1}{2}$ & \hspace{2mm}  $\frac{1}{2}^+$, $\frac{3}{2}^+$ 
& \hspace{4mm} 4273 & -$\frac{\sqrt{3}}{4}$ & $\frac{1}{4}$  & $\frac{3}{4 \sqrt{2}}$ & $ - \frac{3}{4 \sqrt{6}}$ 
 \\[1.1ex]
$P_c(1/2^-)$  & \hspace{1mm} $\frac{1}{2}$ & \hspace{2mm}  $\frac{1}{2}^-$ \hspace{1mm} & \hspace{4mm} 4487
& $\frac{3}{4\sqrt{6}}$ & -$\frac{1}{4\sqrt{2}}$  & -$\frac{1}{8}$   & $\frac{1}{8\sqrt{3}}$ 
 \\[1.1ex]
\hline
\end{tabular}
\end{table}

\section{Conclusions}

The present study relies on  a compact pentaquark picture  of the $uudc \bar c$ pentaquark based on the flavor-spin model extended to SU(4).
An important feature  is that the model introduces an isospin dependence of pentaquarks, necessary to discriminate 
between decay channels.
 
The spin part of the pentaquark state is a mixed state of spin 0 and spin 1 mesons. 
Thus spin 0 and spin 1 mesons can be produced in the decay
of the pentaquark. 
The ratio of the flavor-spin overlap between the pentaquark and the exit channel states 
depends on the recoupling coefficient in the spin part of the pentaquark wave function, Appendix \ref{spinpart},
and thus it depends only on the meson spin when the exit baryon is fixed.
The light quark masses used in the model were adjusted to reproduce the 
experimental values of the involved baryons. The charm quark mass reproduces the 
average mass of  $D$ mesons.  Accordingly, the $J/\psi$ and $\eta_c$ mesons are degenerate, like in the 
heavy-quark spin symmetry limit used in baryon-meson molecular scenario  \cite{Voloshin:2019aut}. 
When the kinematical difference between two final channels, for example, $\eta_c p$ and $J/\psi p$,
are neglected, we obtain for the ratio of the decay rates of either $P_c(1/2^+)$ and $P_c(1/2^-)$  values which are at variance 
with  those of Ref. \cite{Voloshin:2019aut}.

The present work is a first attempt towards estimating ratios of decay widths in the SU(4) flavor-spin model,
based on the approximation of the flavor-spin part of the transition matrix element between the initial and final states
by the overlap of the corresponding wave functions.
The approximation is satisfactory for describing ratios of decay rates of pentaquarks with the same parity. 
The ratio of decay rates of pentaquarks of different parities is affected by
the orbital-color part of the transition matrix element between the initial and final states,
not needed in this simple approach. It will be a challenge to estimate the orbital-color part 
because a deeper understanding of the decay mechanism is necessary. Further experimental
information about the parity of pentaquarks would  stimulate more theoretical work.

For positive parity the procedure can be easily extended to the study of excited pentaquark states 
with  $J^P$ values up to 5/2$^+$, allowed by the symmetry of the 
wave functions \cite{Stancu:2019qga}.


\appendix


\section{Baryons}\label{baryons}

For convenience here we reproduce the flavor wave functions of the baryons needed in this study
with phase conventions consistent with the permutation group $S_3$  \cite{Stancu:1991rc}.
For the proton we have 
\begin{eqnarray}\label{proton}
| \phi^{\rho}_p \rangle & = & \frac{1}{\sqrt{2}}~(udu-duu), \nonumber\\
| \phi^{\lambda}_p \rangle & = & -\frac{1}{\sqrt{6}}(udu+duu-2uud),
\end{eqnarray}
for $\Lambda_c$ they are 
\begin{eqnarray}\label{lambda}
| \phi^{\rho}_{\Lambda_c} \rangle & = & \frac{1}{\sqrt{12}}~(2udc-2duc+cdu-cud+ucd-dcu), \nonumber\\
| \phi^{\lambda}_{\Lambda_c} \rangle & = & -\frac{1}{2}~(cud-cdu+ucd-dcu).
\end{eqnarray}

\section{Flavor states of $uudc \bar c$ pentaquarks}\label{flavorpart}

In the following we reproduce the flavor wave functions obtained in Ref. \cite{Ortiz-Pacheco:2018ccl}
for $uudc \bar c$ pentaquarks $P_c$ needed in this study. They form an octet and have isospin 1/2. 

The flavor states of a pentaquark
$ q^4 \bar q$ are basis states of SU(4) irreducible representations
appearing in the decomposition of the direct product of  $[22]$ or $[211]$  of  ${q^4}$
states and the irreducible representation  of $\bar q$, namely $[111]$. 
Using the notation of Ref. \cite{Ortiz-Pacheco:2018ccl}  one has 
\begin{equation}
E: ~  [22]_{q^4} \otimes [111]_{\bar q} =  [331]_{q^4 \bar q} \oplus [3211]_{q^4 \bar q}
\end{equation}
and 
\begin{equation}
F_1: ~~[211]_{q^4} \otimes [111]_{\bar q} = [322]_{q^4 \bar q} \oplus [3211]_{q^4 \bar q} \oplus [2221]_{q^4 \bar q}.
\end{equation}
The flavor part  of the state (\ref{Pcplus}) is obtained from combinations of basis states of $[331]_{q^4 \bar q}$ and $[3211]_{q^4 \bar q}$.
and the flavor part of the state (\ref{Pcminus}) is obtained from combinations of basis states
of $[322]_{q^4 \bar q}$ and $[3211]_{q^4 \bar q}$.

In this way the pentaquark wave functions (\ref{Pcplus}) containing the symmetry $[22]_F$ of the four-quark subsystem become
\begin{eqnarray}\label{Erho}
| \phi^{\rho}_E \rangle & = & - \frac{1}{2 \sqrt{2}} ~(ucud+uduc-cuud-duuc-ucdu-udcu+cudu+ducu) \bar c, 
\end{eqnarray}
\begin{eqnarray}\label{Elambda}
| \phi^{\lambda}_E \rangle & = &  - \frac{1}{2 \sqrt{6}}~(2uucd+2uudc-ucud-uduc-cuud-duuc \nonumber\\
& + & 2cduu+2dcuu-ucdu-udcu-cudu-ducu)\bar c, 
\end{eqnarray}
where the subscripts $\rho$ and $\lambda$ indicate that the quarks 1 and 2 are in an antisymmetric and symmetric
state respectively.  
Similarly the pentaquark wave functions (\ref{Pcminus}) containing the symmetry $[211]_F$ of the four-quark subsystem are
\begin{eqnarray}\label{F1}
| \phi^{\rho}_{F_1} \rangle & = & - \frac{1}{4 \sqrt{3}} ~(2dcuu-2cduu-udcu+ducu+ucdu-cudu \nonumber \\
& + & 3uduc-3duuc-3ucud+3cuud)\bar c, \\
| \phi^{\lambda}_{F_1} \rangle & = & - \frac{1}{4} ~(2uudc-uduc-duuc-2uucd \nonumber\\
& + & ucud+cuud+udcu+ducu-ucdu-cudu)\bar c, \\
| \phi^{\eta}_{F_1} \rangle & = & - \frac{1}{\sqrt{6}} ~(dcuu-cduu-ducu+udcu+cudu-ucdu) \bar c, 
\end{eqnarray}
corresponding to the three basis states of $[211]_F$. 

\section{The spin part}\label{spinpart}

\appendix


\section{Baryons}\label{baryons}

Here we write the spin wave functions ${\chi}_{SM}$ of $q^4 \bar q$ 
in terms of products of baryon and meson wave functions, needed in the calculation the flavor-spin overlap $O^{FS}$. 
For $S = 1/2$ they are the basis vectors of the irreducible representation  $[32]$ of 
SU(2). 
The first step is to couple the antiquark of spin 1/2 to $q^4$ of spin $s_q $. 
One has 
\begin{equation}
{\chi}_{SM} = \sum_{m_q,m_{\bar q}}  C^{s_q ~1/2 ~S}_{m_q ~m_{\bar q ~M}} ~\chi^{[f]}_{s_q~m_q}~ \chi^{[1]}_{1/2~m_{\bar q}},
\end{equation}
where  $[f]$ and $[1]$ stand for the SU(2) irreducible representations associated to $q^4$ and $\bar q$ respectively. 
For the states (\ref{Pcplus}) and (\ref{Pcminus}) we have $[f]$ = $[22]$, thus $s_q$ = 0.

The following step is to decouple one quark $q$ from $q^4$ which amounts to write
\begin{equation}
\chi^{[f]}_{s_q~m_q} = \sum_{m_1,m_2}   C^{s_1 ~1/2 ~s_q}_{m_1 ~m_2 ~m_q}   \chi^{[f_1]}_{s_1~m_1} \chi^{[1]}_{1/2~m_2},
\end{equation}
which is a linear combination of $q^3$ and $q$ states described by $[f_1]$ and $[1]$ respectively.
Then one has to couple $q$ to $\bar q$ to form a meson of a given spin $s$ = 0 or 1
\begin{equation}
\chi^{[1]}_{1/2~m_2} \chi^{[1]}_{1/2~m_{\bar q}} = \sum_{s,m_s}  C^{1/2 ~1/2 ~s}_{m_2 ~m_{\bar q} ~m_s}~ \chi^{[f']}_{s~m_s},
\end{equation}
where $[f'] = [11]$, $s = 0$ for scalar and  $[f'] = [2]$, $s$ = 1 for vector mesons respectively.
Combining together the coupling and the decoupling one obtains
\begin{equation}
{\chi}_{SM} = \sum_s [(2 s + 1)(2 s_q +1) W(s_1 ~1/2 ~S;~s_q~ s) ( \chi^{[f_1]}_{s_1 ~m_1} \chi^{[f']}_{s~m_s} )^{[32]}_{SM},
\end{equation}
in terms of the Racah coefficient $W(s_1 ~1/2 ~S;~s_q~ s)$. Note that the baryon  $\chi^{[f]}_{s_1 ~m_1}$
and the meson $\chi^{[f']}_{s~m_s}$ states are coupled together to a given spin and projection $SM$.
Here we have $[f_1]$ = $[21]$ thus octet baryons with $s_1$ =1/2. Implementing the Racah coefficient for the two possible values of the 
meson state, $s$ = 0 and $s$ = 1 we obtain the pentaquark spin state as
\begin{equation}\label{pentaspin}
{\chi}_{1/2M} = - \frac{1}{2} ( \chi^{[21]}_{1/2 M} ~\chi^{[11]}_{00} )^{[32]}_{1/2M}  
+ \frac{\sqrt{3}}{2} ( \chi^{[21]}_{1/2 m_1} ~\chi^{[2]}_{1m_s} )^{[32]}_{1/2M},  
\end{equation}
which is a linear combination containing  scalar and vector meson states. 
There are two independent five-quark spin states for which the 
above formula applies. Each is  defined by a Young tableau.
Thus the left hand side of Eq. (\ref{pentaspin}) can be associated  
\begin{equation}
\mbox{ either to}~~~
\raisebox{-15.0pt}{\mbox{\begin{Young}
1 & 2 & 5\cr
3 & 4 \cr
\end{Young}}}~~~\mbox{ or to}~~~
\raisebox{-15.0pt}{\mbox{\begin{Young}
1 & 3 & 5\cr
2 & 4 \cr
\end{Young}}}~,
\end{equation}
corresponding to $\chi^{\lambda}_5$ and $\chi^{\rho}_5$ respectively, introduced in Eq. (\ref{OFSPLUS}).

\section{Flavor-spin wave function for negative parity states}\label{negative}

The flavor-spin parts of the negative parity state 
denoted by $(\phi_{F_{1}} \chi_{1/2M})_i$ ($i$ = 1,2,3) in Eq.(\ref{FSminus}) 
can be obtained by  writing  the spin-flavor state
for the subsystem of four quarks followed by the coupling of the antiquark.
Here we present only the case $i$ =3, the only one which is needed.
The $q^4$ flavor-spin state can be written  in terms of its spin and flavor parts with the
help of isoscalar factors of the permutation group $S_4$ \cite{Stancu:1991rc,Stancu:1999qr}.
After the coupling of the antiquark one obtains
\begin{equation}
(\phi_{F_{1}} \chi_{1/2M})_3
=  -\frac{1}{\sqrt{2}} 
\raisebox{-15.0pt}{\mbox{\begin{Young}
1 & 2 & 5\cr
3 & 4 \cr
\end{Young}}} ~\phi^{\lambda}_{F_{1}}
-\frac{1}{\sqrt{2}} 
\raisebox{-15.0pt}{\mbox{\begin{Young}
1 & 3 & 5\cr
2 & 4 \cr
\end{Young}}} ~\phi^{\rho}_{F_{1}}~,
\end{equation}
where the Young tableaux correspond to spin basis vectors of the pentaquark.
\vspace{1cm}

\acknowledgements

This work has been supported by the Fonds de la Recherche Scientifique - FNRS, Belgium, 
under the grant number 4.4503.19.


 \end{document}